\documentclass[fleqn,usenatbib]{mnras}

\usepackage{newtxtext,newtxmath}
\usepackage[T1]{fontenc}

\DeclareRobustCommand{\VAN}[3]{#2}
\let\VANthebibliography\thebibliography
\def\thebibliography{\DeclareRobustCommand{\VAN}[3]{##3}\VANthebibliography}

\usepackage{graphicx}	% Including figure files
\usepackage{amsmath}	% Advanced maths commands

\graphicspath{{./figs/}{./png/}}

\newcommand{\Fig}[1]{Figure~\ref{#1}}

\newcommand{\App}[1]{Appendix~\ref{#1}}

\newcommand{\we}{Waldmeier Effect}

\newcommand{\msh}{$\mu$Hem}

\title[Variation of sunspot area]{Variation of the sunspot area during the rising and declining phases of the solar cycle supports the toroidal flux loss due to flux emergence}

\author[Bidya Binay Karak et al.]{
Bidya Binay Karak,$^{1}$\thanks{E-mail: karak.phy@iitbhu.ac.in (BBK)}
Soumya Mishra,$^{2}$
Anu Sreedevi$^{1}$
\\
$^{1}$Department of Physics, Indian Institute of Technology (Banaras Hindu University), Varanasi 221005, India\\
$^{2}$School of Physical Sciences, National Institute of Science Education and Research Bhubaneswar, Jatni, Khurda 752050, India
}

\date{Accepted XXX. Received YYY; in original form ZZZ}

\pubyear{\the\year{}}

\begin{document}
\label{firstpage}
\pagerange{\pageref{firstpage}--\pageref{lastpage}}
\maketitle

\begin{abstract}
Sunspots are obvious observable manifestations of the toroidal magnetic field generated through the dynamo in the convection zone. They appear in different sizes, having a wide distribution in their area. We analyse the sunspot group area of the past 13 cycles and the Bipolar Magnetic Region (BMR) flux for Cycles 23 and 24 to explore their area and flux distributions and connect with the theory. We find that, in general, the group area and BMR flux are statistically larger in the rising phase than in the declining phase of the solar cycle. This implies that the rising phase of the solar cycle is prone to drive more intense space weather. We further show that the mean and median of the area distribution during the rising phase are dependent on cycle strength. However, the distribution mean and median are cycle strength-independent or weakly dependent during the decline phases of the solar cycles, particularly during the last three years when the latitudinal bands of all cycles migrate towards the equator along the same trajectory. These results support the theoretical model of nonlinear flux loss due to flux emergence, which explains why solar cycles rise differently but decay similarly.  
\end{abstract}


%%%%%%%%%%%%%%%%%%%%%%%%%%%%%%%%%%%%%%%%%%%%%%%%%%

%%%%%%%%%%%%%%%%% BODY OF PAPER %%%%%%%%%%%%%%%%%%

\section{Introduction}
     \label{Intro} 

Although the amplitude of the solar cycle varies irregularly from one cycle to another, its amplitude can be predicted if the cycle has progressed a few years after the minimum by computing its rise rate \citep{Kane08, Kumar22}. This is due to the existence of a robust relationship, popularly known as the Waldmeier effect \citep{wald}. The classical form of this effect states that a strong cycle takes less time to rise, while a weaker one takes longer \citep{Hat02}.  Due to irregular variations of solar activity within a cycle (short-term fluctuations like quasi-biannual oscillations, and double peaks), it becomes inaccurate to determine the exact times of solar minimum and maximum (and thus the rise time) 
and the cycle amplitude, making it sometimes difficult to establish the \we\ in the solar cycle \citep{Dik08, garg19}. However, if the rise rate instead of the rise time is taken, then the \we\ is robustly reproduced in all proxies of 
the solar cycle data \citep{CS08, KC11}.  

\citet{W55} studied \we\ in a slightly different way. He analyzed the migration of the latitudinal distribution of sunspots in each cycle. He showed that as the activity level (e.g., number of sunspots) increases, 
the parameters (centres and widths) of the butterfly wings of different cycles 
evolve at different rates as the sunspot latitudes migrate towards the equator. 
However, when the activity level declines, the parameters of the butterfly wing evolve in the same way for all cycles \citep[see Figs. 3-4 of][]{CS16}. 

\citet{BKC22} explained this feature using a nonlinear loss of toroidal flux due to magnetic buoyancy 
in the dynamo model. They employed the algorithm of \citet{NC01, NC02} to capture the toroidal 
flux from the base of the convection zone (CZ) to the surface in the form of sunspots in the following way: 
If the toroidal field in any grid above the base of  
CZ exceeds a certain threshold $Bc$ (which was determined to be $10^4$~G), then a fraction of the flux is emerged by simply reducing this flux at that grid point and by adding the same flux on the surface. 
Hence, two processes of the toroidal flux determine the cycle phase: the generation/supply and the loss through flux eruptions. In the rising phase of a cycle, the generation dominates over the loss, keeping the toroidal field well above $Bc$. This results in a rapid increase in the sunspot number in the early phase. However, rapid spot eruption increases the flux loss, and at some point the flux loss dominates the generation, reducing the field close to $Bc$.  The cycle no longer grows. Any new generation of flux is compensated by eruptions, and the model continues to produce spots as the cycle decays, maintaining the field near $Bc$. 
In a strong cycle, the toroidal field is high, and spots erupt more frequently. However, this frequent spot eruption rapidly reduces the toroidal field to near $Bc$, and the cycle begins to decline early.
For a weak cycle, the situation is different. 
It produces spots less frequently, and the activity level grows slowly.
The toroidal flux loss is lower, and the cycle grows slowly for a long time; by then, the activity level moves a bit closer to the equator (compared to a strong cycle, which begins to decline when the activity level is already at high latitude). Eventually, when a sufficient amount of flux is lost, the toroidal field becomes comparable to $Bc$, and the weak cycle also begins to decline at the same rate as the strong cycle.

If the above explanation, as detailed in \citet{BKC22}, holds in the Sun, we expect the sunspot properties, such as magnetic field and flux, in the declining phase  
to be of similar strength, while in the rising phase, they will be cycle dependent (stronger cycles are expected to have bigger spots).
%to be different in the rise and decline phases and the rising phase only to be cycle strength dependent. 
However, the magnetic field observed on the solar cycle has weak or no resemble to the field that had erupted from the deep CZ \citep[due to various processes during the rise 
through the surface layer;][]{Cheung14}. Furthermore, the magnetic field information of the sunspot is not reliable due to measurement issues before 1996 and is limited to only the last three cycles from space based observations \citep{UE2002, SB1995, Watson11, SS2012}. Therefore we should not consider the magnetic field to check the solar cycle phase dependence of the sunspot properties to connect this theory. The sunspot flux for which we have the systematic records from space-based data (from the Michelson Doppler Imager (MDI) on board the Solar and Heliospheric Observatory (SOHO) and Helioseismic and Magnetic Imager (HMI) on board the Solar Dynamic Observatory (SDO)) are available for the last two solar cycles. 
However, the sunspot area---a good proxy of the flux of the sunspot \citep{SH94,ST23}---can be resembled to the magnetic flux of the sunspot forming flux tubes\footnote{While there is no direct theoretical or observational proof of this assumption, several independent studies support this reasoning  \citep[e.g.,][]{FFH95, WFM13, Cameron18, CJ19, CH26}.}, and the records of the sunspot area have been available for several cycles. Fortunately, \citet{Mandal2020} cross-calibrated the sunspot records from various observatories and produced homogeneous data for the last 13 solar cycles.
Therefore, we shall consider the sunspot area during the different phases of the solar cycle
and check if there is any variation in different phases, to support the theory of the solar cycle
as given in \citet{BKC22}. Our results will also point out that the two phases of the solar cycle, namely the rising and declining phases, do not contribute equally to driving the space weather.

\section{Data and Methodology}
\label{sec:method}
We analyze the cross-calibrated and homogeneous catalog of individual corrected sunspot group areas spanning 1874--2025 compiled by \citet{Mandal2020}, which combines measurements from the Royal Greenwich Observatory (RGO), Kislovodsk, Pulkovo, and Debrecen observatories. 
%To separate the rising and declining phases of each solar cycle, we follow the methodology described in \citet{CS16} and \citet{MKB17}. 
Since sunspot group catalogs do not contain direct magnetic information, we use 
%magnetic-field–based data for periods where space-borne magnetograms are available (Cycles 23 and 24). 
space-borne magnetograms from MDI/SOHO (for Cycle 23) and HMI/SDO (Cycle 24). 
For this purpose, we employ the catalog of Automatic Tracking Algorithm for Bipolar Magnetic Regions \citep[AutoTAB;][]{Sreedevi23},
which 
%AutoTAB is a state-of-the-art algorithm designed to 
automatically detects and tracks the evolution of bipolar magnetic structures throughout their near-side lifetime. 
%The AutoTAB catalog contains information for approximately 12,000 tracked BMRs. 
For each BMR, a representative flux value is assigned as the mean of the top 80\% of its maximum flux during its evolution. 
%For Cycle 23, the algorithm utilizes line-of-sight (LOS) magnetograms from MDI/SOHO, while for Cycle 24 it uses LOS magnetograms from the HMI/SDO. 
%The tracking is performed by identifying spatial overlap of BMRs between successive observations, allowing consistent association of regions across time. This approach results in a comprehensive and homogeneous catalog of tracked BMRs spanning multiple solar cycles.

\begin{figure}%[H]
\centering
\includegraphics[width=0.50\textwidth]{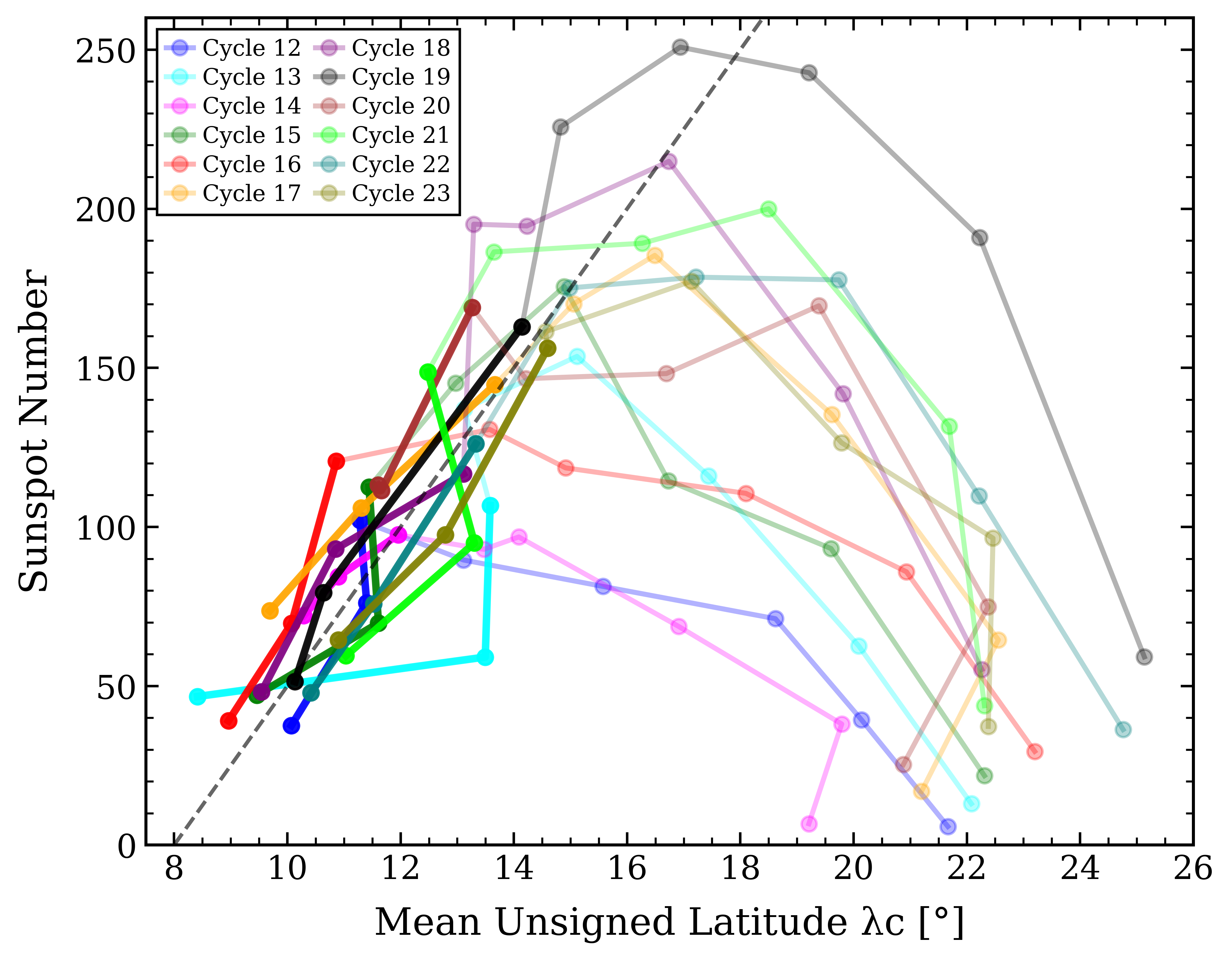}
\caption{The mean of the absolute latitudes of sunspot groups in each year ($\lambda_c$) verses the number of sunspot groups in that year. Different curves are for different cycles. One year data at the beginning and at the end of each cycle are excluded in this analysis. Note that a cycle begins with a high value of $\lambda_c$ on the right side of the plot and with the progress of a cycle when sunspot latitude band migrates equatorward, the trajectory moves towards the left.  The dashed line (having sunspot number $= 25 \lambda_c - 200$) guides the uniform decay of all cycles. The dark parts of the curves highlight the last three years of each cycle during which all cycles fall at the same rate.     
}
\label{fig:cs}
\end{figure}

Individual cycles are identified from one sunspot minimum to the next using the yearly International Sunspot Number (ISN) Version~2.0 provided by WDC--SILSO, Royal Observatory of Belgium. To minimize contamination due to cycle overlap, one year at the beginning and one year at the end of each cycle are excluded from the analysis.

For each cycle, we compute the yearly sunspot number and the mean unsigned latitude of sunspot groups. The variation of these quantities for all cycles is shown in \Fig{fig:cs}. 
Let $S(t)$ denotes the annualized sunspot number. For a given cycle $k$ with start year $t_0^{(k)}$ and end year $t^{(k)}_E$, we define the transition year $t_R^{(k)}$ as the year within the interval $[t_0^{(k)},\, t^{(k)}]$ for which $S(t)$ attains its maximum. The rising phase is defined as the interval from $t_0^{(k)}$ to $t_R^{(k)}-1$, while the declining phase extends from $t_R^{(k)}+1$ to $t^{(k)}$. The transition year itself is treated as a discrete boundary and is excluded from both phases.

For most cycles, the phase boundaries determined using the yearly ISN are consistent with those inferred independently from the evolution of sunspot area and mean latitude in \Fig{fig:cs}. 
%We use yearly mean ISN to define the transition year provides a global and cycle-integrated proxy of solar activity and reduces sensitivity to short-term fluctuations and hemispheric asymmetries. 
In several cycles (notably Cycles 14 and 22), the activity exhibits double peak around the solar maximum \citep{KMB18}. In such cases, we adopt the year corresponding to the latter maximum of the yearly mean sunspot number as $t_R^{(k)}$. This choice reflects the sustained high activity level following the first peak and avoids an artificially early termination of the rising phase. This consideration is particularly important for strong cycles (notably Cycles~18, 19, 21), which are characterized by a rapid early rise followed by an extended period of elevated activity near maximum. Selecting the later peak as $t_R^{(k)}$ ensures that the rising-phase statistics are not biased by the unusually fast initial growth of strong cycles and yields a physically consistent separation between the rising and declining phases across cycles of different strengths. 

%For most cycles, the phase boundaries determined using the yearly mean ISN are consistent with those inferred independently from the evolution of sunspot area and mean latitude in \Fig{fig:cs}. Cycle~14 represents a special case, as $S(t)$ shows two nearly equal maxima corresponding to mean latitudes of approximately $14^\circ$ and $12^\circ$. Either of these peaks could reasonably be adopted as the transition year. In this work, we choose the second peak, which corresponds to the lower mean latitude and is consistent with the equatorward migration of the activity belt and with the transition inferred from the ISN. This definition is used throughout the analysis, except in \Fig{fig:rd_strength}, where the first peak is adopted to maintain consistency with the specific comparison presented in that figure.

%All rising and declining-phase statistics reported in this paper are computed using the phase definitions described above.

\begin{figure}%[H]
\centering
\includegraphics[width=0.49\textwidth]{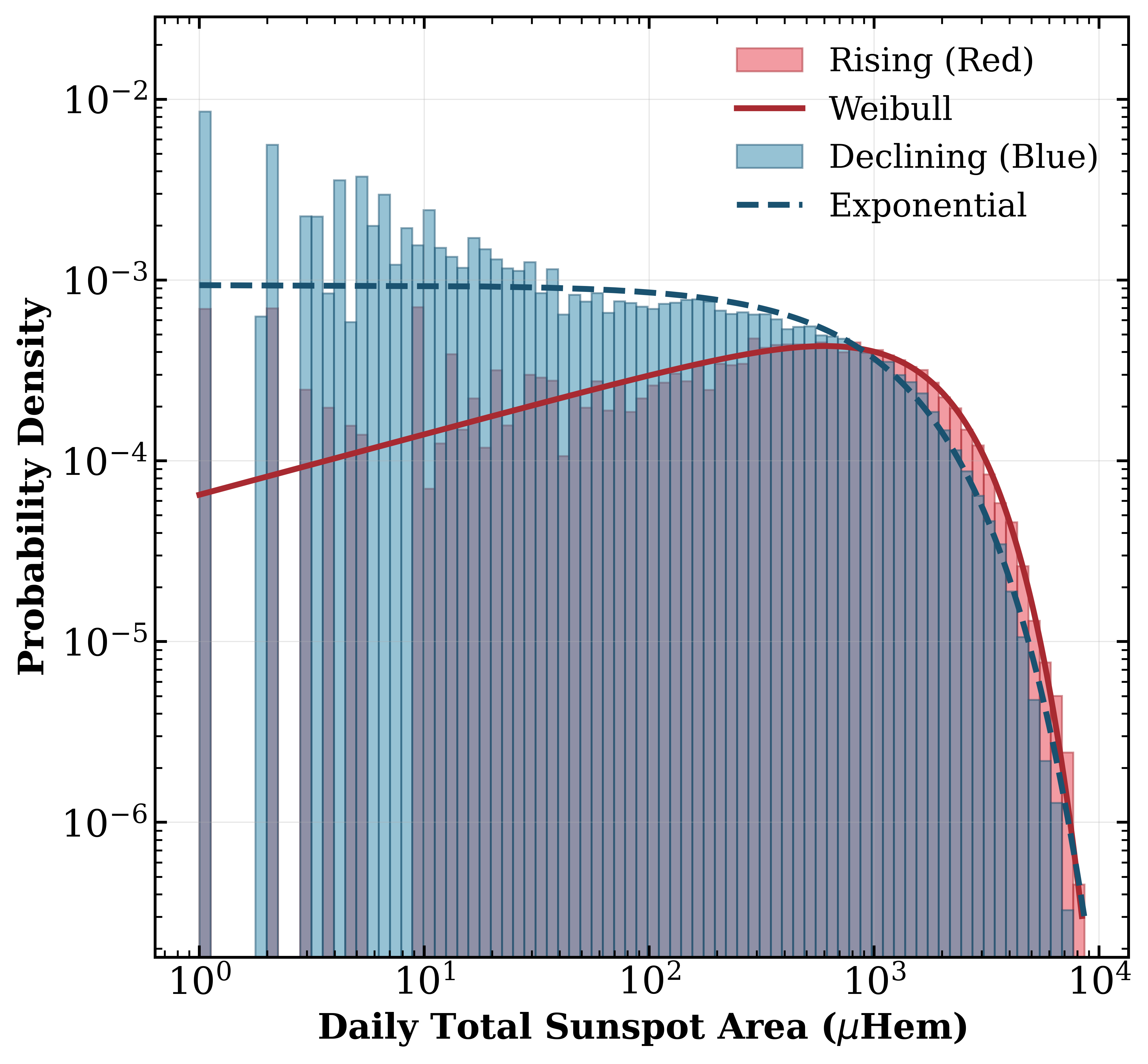}
\caption{Distributions of 
%(a) the individual and (b) 
the daily total sunspot areas (in $\mu\mathrm{Hem}$) during the rising (red) and declining (blue) phases of the solar cycle. 
%The individual sunspot areas are well described by log-normal distributions, with best-fit parameters $\mu = 4.325 \pm 0.005$, $\sigma = 1.525 \pm 0.003$ (rising) and $\mu = 4.272 \pm 0.004$, $\sigma = 1.547 \pm 0.003$ (declining). For daily total sunspot areas, 
The rising-phase distribution is best fitted by a Weibull function ($k = 1.335 \pm 0.010$, $\lambda = 1703.14 \pm 12.42\,\mu\mathrm{Hem}$), 
while the declining-phase distribution is better described by an exponential model ($\lambda = (9.36 \pm 0.06) \times 10^{-4}\,\mu\mathrm{Hem}^{-1}$).}
 \label{fig:dist}
\end{figure}

\begin{figure}%[H]
\centering
\includegraphics[width=0.48\textwidth]{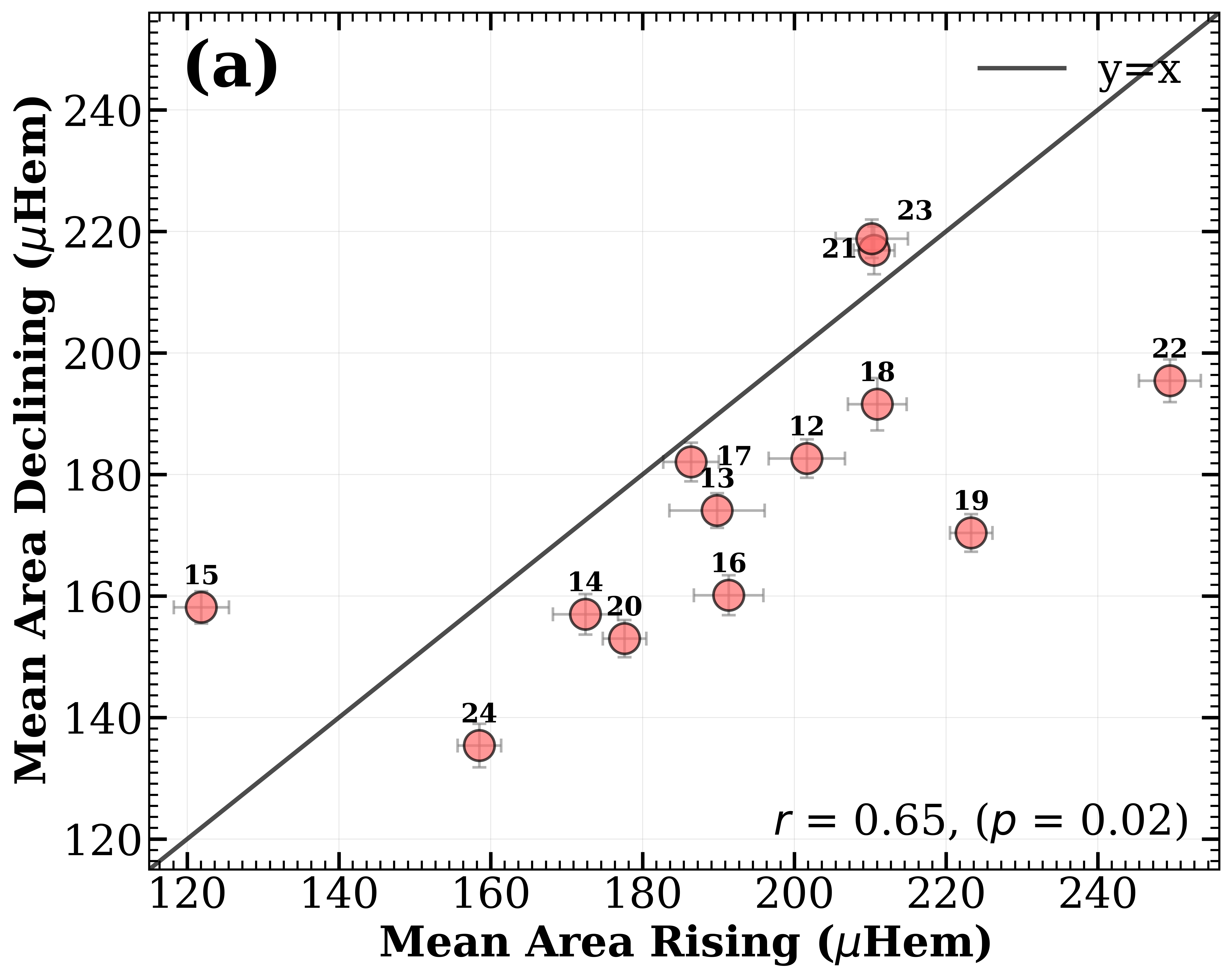}
\includegraphics[width=0.48\textwidth]{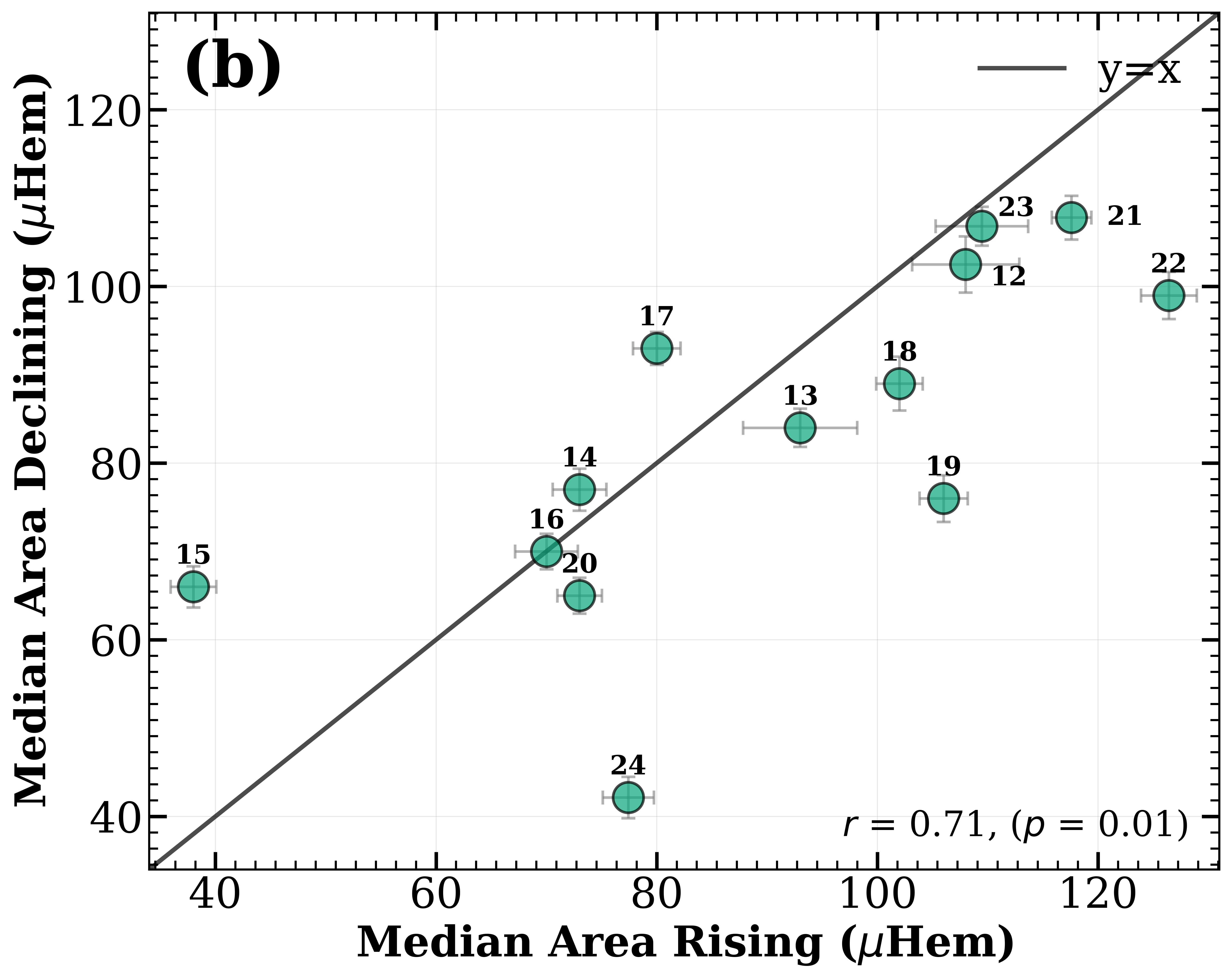}
\caption{Scatter plots of the (a) mean and (b) median area of sunspot groups computed over the rising phase verses the same over the decline phase. Cycle numbers are assigned to each data points. The linear Pearson correlation coefficients ($r$) and $p$ values are printed on each subplot. 
}
\label{fig:lincorrRD}
\end{figure}
\begin{figure*}%[H]
\centering
\includegraphics[width=0.96\textwidth]{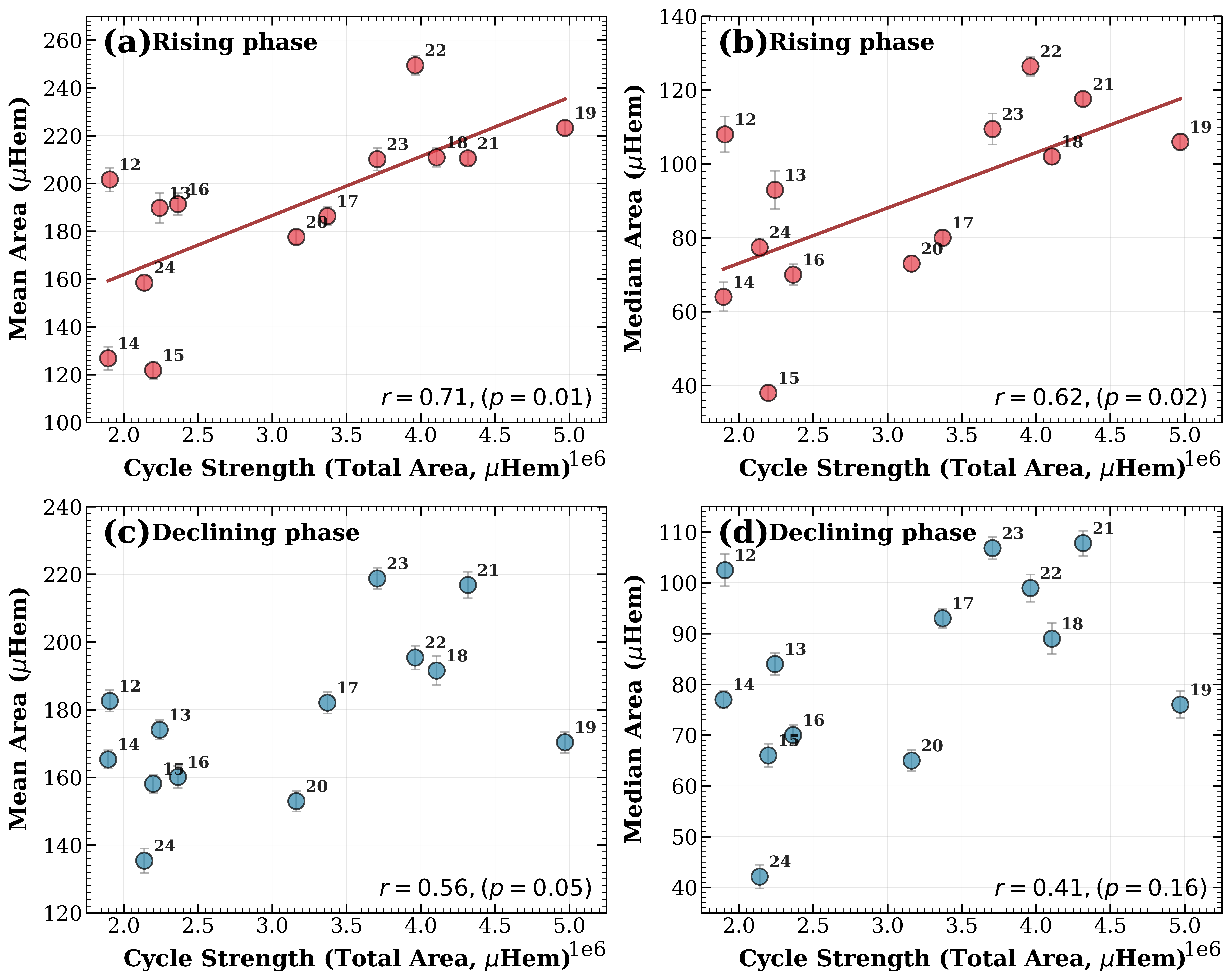}

\caption{Top two panels: Mean area of sunspot groups during the (a) rising phase and (b) declining phase versus the cycle strength (measured by the total group area in \msh) for Solar Cycles 12–24. Points are labelled by the cycle number. The dashed line represents the best linear fit to the data. 
The bottom two panels are the same as the top ones, but for the median area.
The linear Pearson correlation value and $p$ are printed on each panel. 
The Spearman rank correlation coefficients for the four panels are 0.71 ($p = 0.01$), 0.62 ($p = 0.02$), 0.56 ($p = 0.05$), and 0.41 ($p = 0.16$), respectively. }
\label{fig:rd_strength}
\end{figure*}

\section{Results and Discussion}
\label{sec:results}
Isolating the sunspots for the rising and declining phases of all past 13 cycles, we study their properties. In \Fig{fig:dist} we show the distribution of all sunspot group areas during the rising and declining phases, combined for all cycles. Here, we present the distribution from the daily total area data instead of the individual group area, as the latter distribution is somewhat noisy, and the two profiles are not distinctly visible. 
%In this figure we find that the sunspot area both during the rising and decline phases closely follow log-normal distributions and the two distributions appear almost identical. 
%However, when we observe the distributions of the daily (total) group area instead of the individual ones, as shown 
In \Fig{fig:dist}, we clearly identify two distinct distributions corresponding to the rising and declining phases. 
The rising phase shows somewhat narrow distribution with peak at higher area value, while the distribution during the decline phase is peaking at a slightly lower area and has extended tail towards the lower value.
The distribution corresponding to the declining phase is not described well by the log-normal but by the Weibull distribution. These behaviours clearly suggest that the rising phase of the solar cycle is distinct from the decline phase. This we explore in detail below by analysing the distribution properties separately for all the cycles. 

The mean and median values of the area of all individual sunspot groups during the declining phases versus rising phases, computed separately for Cycles 12--24, are shown in \Fig{fig:lincorrRD}.  Several inferences can be made from this plot. For most of the cycles (except Cycles 15, 21, and 23\footnote{For this cycle, sunspot properties extracted from the SOHO/MDI data also do not show significantly lower values in the declining phase than those in the rising phase \citep{Watson11}.} for the mean and Cycles 14--17 for median), the mean or median value of the sunspot area during the rising phase is higher than that of the decline phase. This is consistent with the results presented in the area distribution in two phases, combined for all cycles (\Fig{fig:dist}; also see the distributions of the daily sunspot area in \Fig{fig:distinvidual} for a few cycles). This clearly implies that the Sun produces (statistically) bigger spots during the rising phase than in the decline phase. Observations indeed find that the intensity of the solar flare eruptions increases with the group area, although the complexity of the magnetic structure plays a key role \citep{Sammis00, Ternullo06, Lee12, Watari22, Lin23, rajkumar25}. 
Consequently, the rising phases are prone to drive more violent space weather. Another point is that the data points are not strongly correlated in this plot. This implies that if the sunspot area during the rising phase is large, then it is not necessary that the spot area during the decline phase will also be large. For example, Cycle 19 (which is one of the strongest cycles) has a considerably large value of the mean/median sunspot area during the rising phase, compared to its decline phase. This asymmetry in the sunspot area is high in the strong cycles and less in weak cycles (e.g., Cycles 12--16).  Finally, the mean (or median) area during the rising phase (horizontal axis of this plot) have a larger variation than the area during the decline phase (the vertical axis). This suggests that the sunspot area during the decline phase is less dependent on the  cycle strength. 

\begin{figure*}%[H]
\centering
\includegraphics[width=0.96\textwidth]{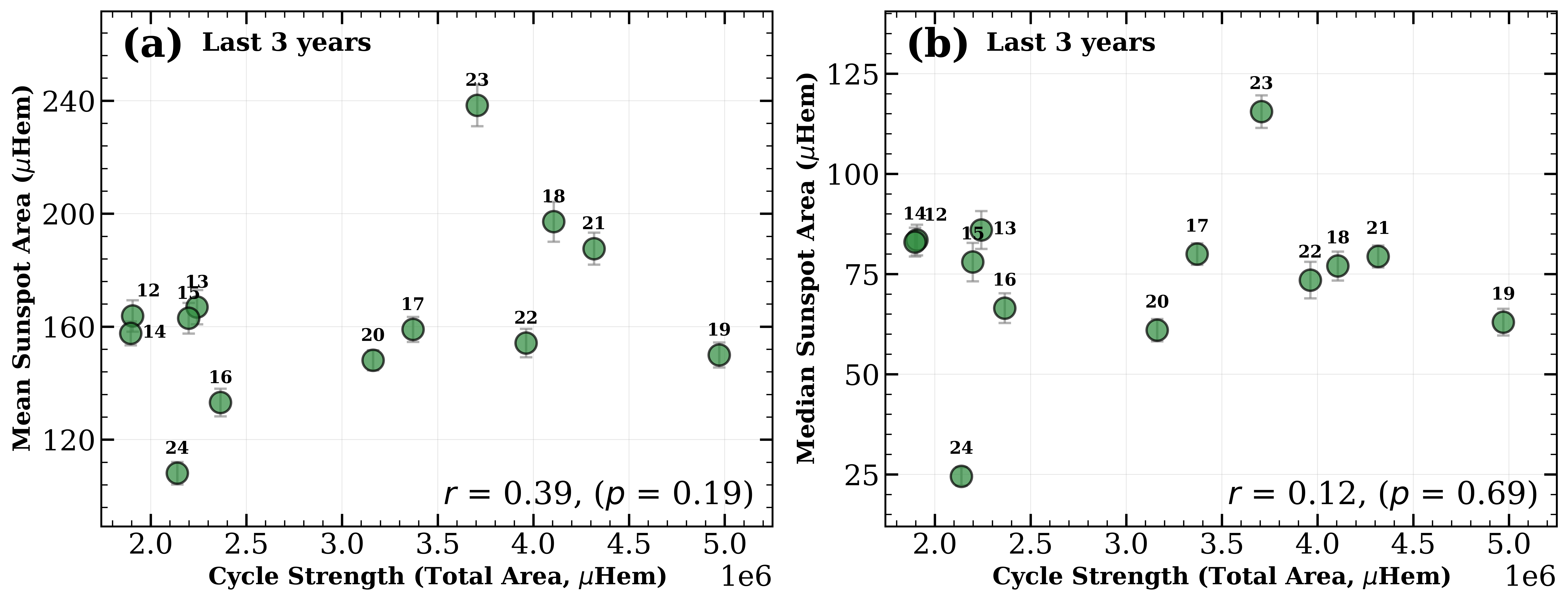}
\caption{The same as the bottom panels of \Fig{fig:rd_strength} but from the sunspot areas during the last 3~years (excluding the final year) of the decline phase of the cycle. 
}
\label{fig:corr3yr}
\end{figure*}

To further explore this feature, we show the scatter plots of the mean and median areas during the rising and falling phases with the cycle strength in \Fig{fig:rd_strength}. Here, the cycle strength is measured as the total area covered by the sunspot groups in the cycle. Instead of total area, if we consider the peak ISN sunspot number as the cycle strength, then also we get similar results.    
We note that in general the (mean and median) sunspot area is correlated with the cycle strength, suggesting that if a cycle is strong, it produces (statistically) bigger spots, consistent with the results of sunspots from the Kodaikanal Solar Observatory \citep{MB16}. This is also the reason why strong cycles are more prone to produce extreme space weather events \citep{chapman20, Owens21}.   However, the correlation between the sunspot area (mean and median) with the cycle strength during the declining phase is less compared to that of during the rising phase.\footnote{If we include the transition year in the rising phase of the cycle, then the correlations in panels (a) and (b) of \Fig{fig:rd_strength} improve slightly.}  Again, this hints towards the fact that the decline phase is less correlated with the cycle strength. 

This feature becomes more transparent when we consider only the sunspot area of the last three years (again we exclude one year at the end). In \Fig{fig:corr3yr}, we observe that the mean and median sunspot areas over the last three years of the cycle are not correlated with the strength of the cycle. We find consistently similar results with the RGO (Greenwich - USAF/NOAA) data as well (\App{app:A2}).

Recalling the theory of \citet{BKC22}, we  know that the activity levels of all cycles fall at the same rate, and in particular the strong cycles begin to decline early (when the activity belt is already at high latitudes) and the weak cycles decline later. As highlighted in \cite{CS16} and in our \Fig{fig:cs}, we find that in the last few years of the cycles all the trajectories (of the activity vs latitude plane) are merged into a common trajectory.  Clearly, this is the phase when the toroidal field at the base of the CZ remains comparable in all  cycles and does not depend on the cycle strength. This is what is indicative here in \Fig{fig:corr3yr}. While we are not observing a strictly constant value for the mean and median sunspot area, which we should not because of their statistical variations form a limited number and observational issues, we clearly observe them as cycle uncorrelated and the spread of the vertical-axes of \Fig{fig:corr3yr} is quite small. Although the values for Cycles 23 and 24 are deviated from the average trend, shifting the 3-year window by one year towards the weaker / intense  phase for Cycle 23/24 helps reduce the spread in \Fig{fig:corr3yr}. 

We note that there is an equally promising alternative explanation for the uniform decay of cycles, as given by \citet{CS16} and later explored in a dynamo model by \citet{Talafha22}, based on diffusive cancellation at the equator. However, in this scenario, the decline phase will still be cycle-dependent; strong cycles are expected to have high flux spots even in their decline phase. 
%for strong cycles, the field strength at least in their initial years of the decline phase should be high. 
In observations (\Fig{fig:rd_strength} bottom panels and \Fig{fig:corr3yr}), we find that even strong cycles (e.g., Cycle 19) do not have a higher sunspot area; all cycles have more or less similar sunspot area. Hence, based on our understanding, this diffusive cancellation idea does not seem to be a favourable explanation for the observed properties of solar cycle decay. However, future studies are required to confirm this. 

Finally, as discussed earlier, we also check the cycle phase dependence of sunspots with LOS magnetic observations from AutoTAB for Cycles 23 and 24. 
%For this purpose, we use the AutoTAB catalog. 
During the rising phase of Cycle 23/24, the mean magnetic fluxes are $5.47 \times 10^{23} \pm 1.9 \times 10^{21}$) / ($1.12\times10^{23} \pm 5.08 \times 10^{20}$)\,Mx, while the corresponding median values are $1.5 \times 10^{22}$ / $1.1\times10^{22}$\,Mx. In the declining phase, the mean flux values for Cycle 23/24 are $3.04 \times 10^{23} \pm 8.9 \times 10^{20}$ / $1.21\times10^{22} \pm 1.5\times 10^{19}$\,Mx with median values of $1.34 \times 10^{22}$ / $3.12\times10^{21}$\, Mx, respectively. We note that the AutoTAB catalog includes not only sunspot-forming regions, but also weaker BMRs that do not appear as sunspots, as well as some high-flux ephemeral regions \citep{Sreedevi24}. Interestingly, the higher mean and median flux values observed during the rising phase are reflected not only in sunspot data but also in the broader BMR population, highlighting a consistent magnetic signature across different scales of flux emergence.

\section{Conclusion}
\label{sec:conclusion}

%In this article, 
We have provided pieces of observational evidence behind the theory of \citet{BKC22}, which explained a fundamental feature of the solar cycle: Different cycles rise at different rates while their decline phase is similar \citep{W55, CS16}.  They employed the buoyancy algorithm of 
\citet{NC01, NC02} in solar dynamo, namely, when the toroidal magnetic field in the CZ exceeds a certain threshold, a fraction of the toroidal flux is released from there to form a sunspot on the solar surface.  Using this idea, \citet{BKC22} showed that in the early phase of the solar cycle, the toroidal field is considerably higher than this threshold, allowing the cycle to grow fast. As every sunspot eruption releases some flux, the Sun releases a considerable amount of flux during the early phase of the solar cycle, bringing the field down to the threshold, causing the cycle to decline. This flux loss happens at a faster rate for a strong cycle, causing the cycle to decline early when the activity belt is already at high latitudes (compared to a weak cycle, whose flux loss occurs at a slow rate and the toroidal belt at the base of the CZ is advected towards the equator). Thus, during the rising phase of the solar cycle, we expect large spots, and they should correlate with the cycle strength. In contrast, for the decline phase (at least for the last three years when all cycles decline in the same way), since the field in the CZ is comparable to the threshold, we expect small spots and sunspot properties to be cycle-independent.     
By analysing the areas of the sunspot groups for the last 13 solar cycles and BMRs from LOS magnetograms from the last two cycles, we find general agreement with the expectation from the above theory. In particular, we find that (i) the distribution of sunspot group areas of the rising phase is shifted towards higher values compared to that of the decline phase and (ii) the mean and median values of the sunspot areas during the rising phase are correlated with the cycle strength, while the decline phase of the area distribution is largely uncorrelated with the cycle strength. Our analysis also suggests that the rising phase of a cycle is prone to cause more intense space weather.

%Using the toroidal flux loss in the dynamo theory,  showed that a strong cycle which produced  high toroidal magnetic flux in the CZ loses flux quickly at its early phase and   The predictions of this theory were that the sunspot flux/area should be bigger during the rising phase, and they should be correlated with the cycle strength. However, during the falling phase, the spot area should be less compared to that during the rising phase, and it should be independent of the cycle strength. By analysing the areas of the sunspot groups for the last 13 solar cycles, we find exactly these same properties.

\section*{Acknowledgements}
The authors sincerely thank the  reviewer for offering several constructive comments and questions; answering them helps improve the quality of the article. The authors also are indebted to Robert Cameron for a helpful discussion and suggestion.  B.B.K. acknowledges the financial support from the Anusandhan National Research Foundation (ANRF) through the MATRIC program (file no. MTR/2023/000670).
%%%%%%%%%%%%%%%%%%%%%%%%%%%%%%%%%%%%%%%%%%%%%%%%%%
\section*{Data Availability}
Sunspot area data were obtained from \citet{Mandal2020} who produced cross-calibrated and corrected areas of individual sunspot groups during 1874--2025, 
and from Royal Observatory, Greenwich - USAF/NOAA Sunspot Data. The International Sunspot Number V2.0 is obtained from WDC-SILSO, Royal Observatory of Belgium, Brussels. 
AutoTAB catalog is available at  \url{https://github.com/sreedevi-anu/AutoTAB}
\bibliographystyle{mnras}
\bibliography{ms}  

@ARTICLE{Mandal2020,
       author = {{Mandal}, Sudip and {Krivova}, Natalie A. and {Solanki}, Sami K. and {Sinha}, Nimesh and {Banerjee}, Dipankar},
        title = "{Sunspot area catalog revisited: Daily cross-calibrated areas since 1874}",
      journal = {\aap},
         year = 2020,
        month = aug,
       volume = {640},
          eid = {A78},
        pages = {A78},
          doi = {10.1051/0004-6361/202037547},
archivePrefix = {arXiv},
       eprint = {2004.14618},
 primaryClass = {astro-ph.SR},
       adsurl = {https://ui.adsabs.harvard.edu/abs/2020A&A...640A..78M}
}

@ARTICLE{UE2002,
       author = {{Ulrich}, Roger K. and {Evans}, Scott and {Boyden}, John E. and {Webster}, Larry},
        title = "{Mount Wilson Synoptic Magnetic Fields: Improved Instrumentation, Calibration, and Analysis Applied to the 2000 July 14 Flare and to the Evolution of the Dipole Field}",
      journal = {\apjs},
         year = 2002,
        month = mar,
       volume = {139},
       number = {1},
        pages = {259-279},
          doi = {10.1086/337948},
       adsurl = {https://ui.adsabs.harvard.edu/abs/2002ApJS..139..259U}
}

@ARTICLE{SB1995,
       author = {{Scherrer}, P.~H. and {Bogart}, R.~S. and {Bush}, R.~I. and {Hoeksema}, J.~T. and {Kosovichev}, A.~G. and {Schou}, J. and {Rosenberg}, W. and {Springer}, L. and {Tarbell}, T.~D. and {Title}, A. and {Wolfson}, C.~J. and {Zayer}, I. and {MDI Engineering Team}},
        title = "{The Solar Oscillations Investigation - Michelson Doppler Imager}",
      journal = {\solphys},
         year = 1995,
        month = dec,
       volume = {162},
       number = {1-2},
        pages = {129-188},
          doi = {10.1007/BF00733429},
       adsurl = {https://ui.adsabs.harvard.edu/abs/1995SoPh..162..129S}
}

@ARTICLE{SS2012,
       author = {{Schou}, J. and {Scherrer}, P.~H. and {Bush}, R.~I. and {Wachter}, R. and {Couvidat}, S. and {Rabello-Soares}, M.~C. and {Bogart}, R.~S. and {Hoeksema}, J.~T. and {Liu}, Y. and {Duvall}, T.~L. and {Akin}, D.~J. and {Allard}, B.~A. and {Miles}, J.~W. and {Rairden}, R. and {Shine}, R.~A. and {Tarbell}, T.~D. and {Title}, A.~M. and {Wolfson}, C.~J. and {Elmore}, D.~F. and {Norton}, A.~A. and {Tomczyk}, S.},
        title = "{Design and Ground Calibration of the Helioseismic and Magnetic Imager (HMI) Instrument on the Solar Dynamics Observatory (SDO)}",
      journal = {\solphys},
         year = 2012,
        month = jan,
       volume = {275},
       number = {1-2},
        pages = {229-259},
          doi = {10.1007/s11207-011-9842-2},
       adsurl = {https://ui.adsabs.harvard.edu/abs/2012SoPh..275..229S}
}

@ARTICLE{Sreedevi24,
author = {{Sreedevi}, Anu and {Jha}, Bibhuti Kumar and {Karak}, Bidya Binay and {Banerjee}, Dipankar},
        title = "{Analysis of BMR Tilt from AutoTAB Catalog: Hinting toward the Thin Flux Tube Model?}",
      journal = {\apj},
         year = 2024,
        month = may,
       volume = {966},
       number = {1},
          eid = {112},
        pages = {112},
          doi = {10.3847/1538-4357/ad34b8},
archivePrefix = {arXiv},
       eprint = {2403.09229},
 primaryClass = {astro-ph.SR},
       adsurl = {https://ui.adsabs.harvard.edu/abs/2024ApJ...966..112S}
}

@ARTICLE{Sreedevi23,
       author = {{Sreedevi}, Anu and {Jha}, Bibhuti Kumar and {Karak}, Bidya Binay and {Banerjee}, Dipankar},
        title = "{AutoTAB: Automatic Tracking Algorithm for Bipolar Magnetic Regions}",
      journal = {arXiv e-prints},
         year = 2023,
        month = apr,
          eid = {arXiv:2304.06615},
        pages = {arXiv:2304.06615},
          doi = {10.48550/arXiv.2304.06615},
archivePrefix = {arXiv},
       eprint = {2304.06615},
 primaryClass = {astro-ph.SR},
       adsurl = {https://ui.adsabs.harvard.edu/abs/2023arXiv230406615S}
}

@ARTICLE{FFH95,
       author = {{Fisher}, G.~H. and {Fan}, Y. and {Howard}, R.~F.},
        title = "{Comparisons between Theory and Observation of Active Region Tilts}",
      journal = {\apj},
         year = 1995,
        month = jan,
       volume = {438},
        pages = {463},
          doi = {10.1086/175090},
       adsurl = {https://ui.adsabs.harvard.edu/abs/1995ApJ...438..463F}
}

@ARTICLE{garg19,
       author = {{Garg}, Suyog and {Karak}, Bidya Binay and {Egeland}, Ricky and
         {Soon}, Willie and {Baliunas}, Sallie},
        title = "{Waldmeier Effect in Stellar Cycles}",
      journal = {arXiv e-prints},
         year = "2019",
        month = "Sep",
          eid = {arXiv:1909.12148},
        pages = {arXiv:1909.12148},
archivePrefix = {arXiv},
       eprint = {1909.12148},
 primaryClass = {astro-ph.SR},
       adsurl = {https://ui.adsabs.harvard.edu/abs/2019arXiv190912148G}
}

@ARTICLE{BKC22,
       author = {{Biswas}, Akash and {Karak}, Bidya Binay and {Cameron}, Robert},
        title = "{Toroidal Flux Loss due to Flux Emergence Explains why Solar Cycles Rise Differently but Decay in a Similar Way}",
      journal = {\prl},
         year = 2022,
        month = dec,
       volume = {129},
       number = {24},
          eid = {241102},
        pages = {241102},
          doi = {10.1103/PhysRevLett.129.241102},
archivePrefix = {arXiv},
       eprint = {2210.07061},
 primaryClass = {astro-ph.SR},
       adsurl = {https://ui.adsabs.harvard.edu/abs/2022PhRvL.129x1102B}
}

@ARTICLE{Cameron18,
       author = {{Cameron}, R.~H. and {Duvall}, T.~L. and {Sch{\"u}ssler}, M. and {Schunker}, H.},
        title = "{Observing and modeling the poloidal and toroidal fields of the solar dynamo}",
      journal = {\aap},
         year = 2018,
        month = jan,
       volume = {609},
          eid = {A56},
        pages = {A56},
          doi = {10.1051/0004-6361/201731481},
archivePrefix = {arXiv},
       eprint = {1710.07126},
 primaryClass = {astro-ph.SR},
       adsurl = {https://ui.adsabs.harvard.edu/abs/2018A&A...609A..56C}
}

@ARTICLE{CH26,
       author = {{Chatterjee}, Soumyadeep and {Hazra}, Gopal},
        title = "{Probing the large-scale magnetic field inside the Sun from three decades of observed surface magnetograms}",
      journal = {arXiv e-prints},
         year = 2025,
        month = sep,
          eid = {arXiv:2509.23959},
        pages = {arXiv:2509.23959},
          doi = {10.48550/arXiv.2509.23959},
archivePrefix = {arXiv},
       eprint = {2509.23959},
 primaryClass = {astro-ph.SR},
       adsurl = {https://ui.adsabs.harvard.edu/abs/2025arXiv250923959C}
}

@ARTICLE{CJ19,
       author = {{Cameron}, R.~H. and {Jiang}, J.},
        title = "{The relationship between flux emergence and subsurface toroidal magnetic flux}",
      journal = {\aap},
         year = 2019,
        month = nov,
       volume = {631},
          eid = {A27},
        pages = {A27},
          doi = {10.1051/0004-6361/201834852},
archivePrefix = {arXiv},
       eprint = {1909.06828},
 primaryClass = {astro-ph.SR},
       adsurl = {https://ui.adsabs.harvard.edu/abs/2019A&A...631A..27C}
}

@ARTICLE{chapman20,
       author = {{Chapman}, S.~C. and {Horne}, R.~B. and {Watkins}, N.~W.},
        title = "{Using the aa Index Over the Last 14 Solar Cycles to Characterize Extreme Geomagnetic Activity}",
      journal = {\grl},
         year = 2020,
        month = feb,
       volume = {47},
       number = {3},
          eid = {e86524},
        pages = {e86524},
          doi = {10.1029/2019GL08652410.1002/essoar.10501379.2},
       adsurl = {https://ui.adsabs.harvard.edu/abs/2020GeoRL..4786524C}
}

@ARTICLE{Cheung14,
       author = {{Cheung}, Mark C.~M. and {Isobe}, Hiroaki},
        title = "{Flux Emergence (Theory)}",
      journal = {Living Reviews in Solar Physics},
         year = 2014,
        month = dec,
       volume = {11},
       number = {1},
          eid = {3},
        pages = {3},
          doi = {10.12942/lrsp-2014-3},
       adsurl = {https://ui.adsabs.harvard.edu/abs/2014LRSP...11....3C}
}

@ARTICLE{CS08,
   author = {{Cameron}, R. and {Sch{\"u}ssler}, M.},
    title = "{A Robust Correlation between Growth Rate and Amplitude of Solar Cycles: Consequences for Prediction Methods}",
  journal = {\apj},
     year = 2008,
    month = oct,
   volume = 685,
    pages = {1291-1296},
      doi = {10.1086/591079},
   adsurl = {http://adsabs.harvard.edu/abs/2008ApJ...685.1291C}
}

@ARTICLE{CS16,
   author = {{Cameron}, R.~H. and {Sch{\"u}ssler}, M.},
    title = "{The turbulent diffusion of toroidal magnetic flux as inferred from properties of the sunspot butterfly diagram}",
  journal = {\aap},
archivePrefix = "arXiv",
   eprint = {1604.07340},
 primaryClass = "astro-ph.SR",
     year = 2016,
    month = jun,
   volume = 591,
      eid = {A46},
    pages = {A46},
      doi = {10.1051/0004-6361/201527284},
   adsurl = {http://adsabs.harvard.edu/abs/2016A%26A...591A..46C}
}

@ARTICLE{Dik08,
   author = {{Dikpati}, M. and {Gilman}, P.~A. and {de Toma}, G.},
    title = "{The Waldmeier Effect: An Artifact of the Definition of Wolf Sunspot Number?}",
  journal = {\apjl},
     year = 2008,
    month = jan,
   volume = 673,
    pages = {L99},
      doi = {10.1086/527360},
   adsurl = {http://adsabs.harvard.edu/abs/2008ApJ...673L..99D}
}

@ARTICLE{Hat02,
       author = {{Hathaway}, David H. and {Wilson}, Robert M. and {Reichmann}, Edwin J.},
        title = "{Group Sunspot Numbers: Sunspot Cycle Characteristics}",
      journal = {\solphys},
         year = 2002,
        month = dec,
       volume = {211},
       number = {1},
        pages = {357-370},
          doi = {10.1023/A:1022425402664},
       adsurl = {https://ui.adsabs.harvard.edu/abs/2002SoPh..211..357H}
}

@ARTICLE{rajkumar25,
       author = {{Kumar}, Raj and {Chandra}, Ramesh and {Pande}, Bimal and {Pande}, Seema},
        title = "{Study of solar energetic particles: their source regions, flares and CMEs during solar cycles 23{\textendash}24}",
      journal = {Indian Journal of Physics},
         year = 2025,
        month = sep,
       volume = {99},
       number = {10},
        pages = {3593-3605},
          doi = {10.1007/s12648-025-03619-8},
archivePrefix = {arXiv},
       eprint = {2504.13654},
 primaryClass = {astro-ph.SR},
       adsurl = {https://ui.adsabs.harvard.edu/abs/2025InJPh..99.3593K}
}

@ARTICLE{Kane08,
   author = {{Kane}, R.~P.},
    title = "{How useful is the Waldmeier effect for prediction of a sunspot cycle?}",
  journal = {Journal of Atmospheric and Solar-Terrestrial Physics},
     year = 2008,
    month = aug,
   volume = 70,
    pages = {1533-1540},
      doi = {10.1016/j.jastp.2008.04.010},
   adsurl = {http://adsabs.harvard.edu/abs/2008JASTP..70.1533K}
}

@ARTICLE{KMB18,
       author = {{Karak}, Bidya Binay and {Mandal}, Sudip and {Banerjee}, Dipankar},
        title = "{Double Peaks of the Solar Cycle: An Explanation from a Dynamo Model}",
      journal = {\apj},
         year = 2018,
        month = oct,
       volume = {866},
       number = {1},
          eid = {17},
        pages = {17},
          doi = {10.3847/1538-4357/aada0d},
archivePrefix = {arXiv},
       eprint = {1808.03922},
 primaryClass = {astro-ph.SR},
       adsurl = {https://ui.adsabs.harvard.edu/abs/2018ApJ...866...17K}
}

@ARTICLE{KC11,
   author = {{Karak}, B.~B. and {Choudhuri}, A.~R.},
    title = "{The Waldmeier effect and the flux transport solar dynamo}",
  journal = {\mnras},
archivePrefix = "arXiv",
   eprint = {1008.0824},
 primaryClass = "astro-ph.SR",
     year = 2011,
    month = jan,
   volume = 410,
    pages = {1503-1512},
      doi = {10.1111/j.1365-2966.2010.17531.x},
   adsurl = {http://adsabs.harvard.edu/abs/2011MNRAS.410.1503K}
}

@ARTICLE{Kumar22,
       author = {{Kumar}, Pawan and {Biswas}, Akash and {Karak}, Bidya Binay},
        title = "{Physical link of the polar field buildup with the Waldmeier effect broadens the scope of early solar cycle prediction: Cycle 25 is likely to be slightly stronger than Cycle 24}",
      journal = {\mnras},
         year = 2022,
        month = jun,
       volume = {513},
       number = {1},
        pages = {L112-L116},
          doi = {10.1093/mnrasl/slac043},
archivePrefix = {arXiv},
       eprint = {2203.11494},
 primaryClass = {astro-ph.SR},
       adsurl = {https://ui.adsabs.harvard.edu/abs/2022MNRAS.513L.112K}
}

@ARTICLE{Lin23,
       author = {{Lin}, Jiaqi and {Wang}, Feng and {Deng}, Linhua and {Deng}, Hui and {Mei}, Ying and {Zhang}, Xiaojuan},
        title = "{Evolutionary Relationship between Sunspot Groups and Soft X-Ray Flares over Solar Cycles 21-25}",
      journal = {\apj},
         year = 2023,
        month = nov,
       volume = {958},
       number = {1},
          eid = {1},
        pages = {1},
          doi = {10.3847/1538-4357/ad0469},
       adsurl = {https://ui.adsabs.harvard.edu/abs/2023ApJ...958....1L}
}

@ARTICLE{Lee12,
       author = {{Lee}, Kangjin and {Moon}, Y.-J. and {Lee}, Jin-Yi and {Lee}, Kyoung-Sun and {Na}, Hyeonock},
        title = "{Solar Flare Occurrence Rate and Probability in Terms of the Sunspot Classification Supplemented with Sunspot Area and Its Changes}",
      journal = {\solphys},
         year = 2012,
        month = dec,
       volume = {281},
       number = {2},
        pages = {639-650},
          doi = {10.1007/s11207-012-0091-9},
       adsurl = {https://ui.adsabs.harvard.edu/abs/2012SoPh..281..639L}
}

@ARTICLE{MB16,
       author = {{Mandal}, Sudip and {Banerjee}, Dipankar},
        title = "{Sunspot Sizes and the Solar Cycle: Analysis Using Kodaikanal White-light Digitized Data}",
      journal = {\apjl},
         year = 2016,
        month = oct,
       volume = {830},
       number = {2},
          eid = {L33},
        pages = {L33},
          doi = {10.3847/2041-8205/830/2/L33},
archivePrefix = {arXiv},
       eprint = {1610.02531},
 primaryClass = {astro-ph.SR},
       adsurl = {https://ui.adsabs.harvard.edu/abs/2016ApJ...830L..33M}
}

@ARTICLE{NC01,
       author = {{Nandy}, Dibyendu and {Choudhuri}, Arnab Rai},
        title = "{Toward a Mean Field Formulation of the Babcock-Leighton Type Solar Dynamo. I. {\ensuremath{\alpha}}-Coefficient versus Durney's Double-Ring Approach}",
      journal = {\apj},
         year = 2001,
        month = apr,
       volume = {551},
       number = {1},
        pages = {576-585},
          doi = {10.1086/320057},
archivePrefix = {arXiv},
       eprint = {astro-ph/0107466},
 primaryClass = {astro-ph},
       adsurl = {https://ui.adsabs.harvard.edu/abs/2001ApJ...551..576N}
}

@ARTICLE{NC02,
       author = {{Nandy}, Dibyendu and {Choudhuri}, Arnab Rai},
        title = "{Explaining the Latitudinal Distribution of Sunspots with Deep Meridional Flow}",
      journal = {Science},
         year = 2002,
        month = may,
       volume = {296},
       number = {5573},
        pages = {1671-1673},
          doi = {10.1126/science.1070955},
       adsurl = {https://ui.adsabs.harvard.edu/abs/2002Sci...296.1671N}
}

@ARTICLE{Owens21,
       author = {{Owens}, Mathew J. and {Lockwood}, Mike and {Barnard}, Luke A. and {Scott}, Chris J. and {Haines}, Carl and {Macneil}, Allan},
        title = "{Extreme Space-Weather Events and the Solar Cycle}",
      journal = {\solphys},
         year = 2021,
        month = may,
       volume = {296},
       number = {5},
          eid = {82},
        pages = {82},
          doi = {10.1007/s11207-021-01831-3},
       adsurl = {https://ui.adsabs.harvard.edu/abs/2021SoPh..296...82O}
}

@ARTICLE{ST23,
       author = {{Sakurai}, Takashi and {Toriumi}, Shin},
        title = "{Probability Distribution Functions of Sunspot Magnetic Flux}",
      journal = {\apj},
         year = 2023,
        month = jan,
       volume = {943},
       number = {1},
          eid = {10},
        pages = {10},
          doi = {10.3847/1538-4357/aca28a},
archivePrefix = {arXiv},
       eprint = {2211.13957},
 primaryClass = {astro-ph.SR},
       adsurl = {https://ui.adsabs.harvard.edu/abs/2023ApJ...943...10S}
}

@ARTICLE{Sammis00,
       author = {{Sammis}, Ian and {Tang}, Frances and {Zirin}, Harold},
        title = "{The Dependence of Large Flare Occurrence on the Magnetic Structure of Sunspots}",
      journal = {\apj},
         year = 2000,
        month = sep,
       volume = {540},
       number = {1},
        pages = {583-587},
          doi = {10.1086/309303},
       adsurl = {https://ui.adsabs.harvard.edu/abs/2000ApJ...540..583S}
}

@ARTICLE{SH94,
   author = {{Schrijver}, C.~J. and {Harvey}, K.~L.},
    title = "{The photospheric magnetic flux budget}",
  journal = {\solphys},
     year = 1994,
    month = mar,
   volume = 150,
    pages = {1-18},
      doi = {10.1007/BF00712873},
   adsurl = {http://adsabs.harvard.edu/abs/1994SoPh..150....1S}
}

@ARTICLE{Ternullo06,
       author = {{Ternullo}, M. and {Contarino}, L. and {Romano}, P. and {Zuccarello}, F.},
        title = "{A statistical analysis of sunspot groups hosting M and X flares}",
      journal = {Astronomische Nachrichten},
         year = 2006,
        month = jan,
       volume = {327},
       number = {1},
        pages = {36-43},
          doi = {10.1002/asna.200510485},
       adsurl = {https://ui.adsabs.harvard.edu/abs/2006AN....327...36T}
}

@ARTICLE{Talafha22,
       author = {{Talafha}, M. and {Nagy}, M. and {Lemerle}, A. and {Petrovay}, K.},
        title = "{Role of observable nonlinearities in solar cycle modulation}",
      journal = {\aap},
         year = 2022,
        month = apr,
       volume = {660},
          eid = {A92},
        pages = {A92},
          doi = {10.1051/0004-6361/202142572},
archivePrefix = {arXiv},
       eprint = {2112.14465},
 primaryClass = {astro-ph.SR},
       adsurl = {https://ui.adsabs.harvard.edu/abs/2022A&A...660A..92T}
}

@ARTICLE{Watson11,
       author = {{Watson}, F.~T. and {Fletcher}, L. and {Marshall}, S.},
        title = "{Evolution of sunspot properties during solar cycle 23}",
      journal = {\aap},
         year = 2011,
        month = sep,
       volume = {533},
          eid = {A14},
        pages = {A14},
          doi = {10.1051/0004-6361/201116655},
archivePrefix = {arXiv},
       eprint = {1108.4285},
 primaryClass = {astro-ph.SR},
       adsurl = {https://ui.adsabs.harvard.edu/abs/2011A&A...533A..14W}
}

@ARTICLE{Watari22,
       author = {{Watari}, Shinichi},
        title = "{Extremely large flares/multiple large flares expected from sunspot groups with large area}",
      journal = {Earth, Planets and Space},
         year = 2022,
        month = dec,
       volume = {74},
       number = {1},
          eid = {115},
        pages = {115},
          doi = {10.1186/s40623-022-01676-5},
       adsurl = {https://ui.adsabs.harvard.edu/abs/2022EP&S...74..115W}
}

@ARTICLE{W55,
       author = {{Waldmeier}, Max},
        title = "{Ergebnisse und Probleme der Sonnenforschung.}",
        journal = {Ergebnisse und Probleme der Sonnenforschung (Leipzig:
Geest \& Portig)},
         year = 1955,
       adsurl = {https://ui.adsabs.harvard.edu/abs/1955epds.book.....W}
}

@ARTICLE{wald,
   author = {{Waldmeier}, M.},
    title = "{Neue Eigenschaften der Sonnenfleckenkurve}",
  journal = {Astronomische Mitteilungen der Eidgen{\"o}ssischen Sternwarte Zurich},
     year = 1935,
   volume = 14,
    pages = {105-136},
   adsurl = {http://adsabs.harvard.edu/abs/1935MiZur..14..105W}
}

@ARTICLE{WFM13,
       author = {{Weber}, M.~A. and {Fan}, Y. and {Miesch}, M.~S.},
        title = "{Comparing Simulations of Rising Flux Tubes Through the Solar Convection Zone with Observations of Solar Active Regions: Constraining the Dynamo Field Strength}",
      journal = {\solphys},
         year = "2013",
        month = "Oct",
       volume = {287},
       number = {1-2},
        pages = {239-263},
          doi = {10.1007/s11207-012-0093-7},
archivePrefix = {arXiv},
       eprint = {1208.1292},
 primaryClass = {astro-ph.SR},
       adsurl = {https://ui.adsabs.harvard.edu/abs/2013SoPh..287..239W}
}

\clearpage
%%%%%%%%%%%%%%%%% APPENDICES %%%%%%%%%%%%%%%%%%%%%
\appendix
\section{Supporting materials}

\subsection{Distribution of sunspot area of individual cycles}

As different cycles have different mean and median values of the sunspot area, the combined distribution of all sunspots does not reflect the true properties of the sunspot area for the rising and declining phases of the individual cycles. Therefore, here in \Fig{fig:distinvidual} we  show the distributions of the daily sunspot areas of rising and declining phases from a few representative cycles.

\begin{figure*}%[H]
\centering
\includegraphics[width=0.85\textwidth]{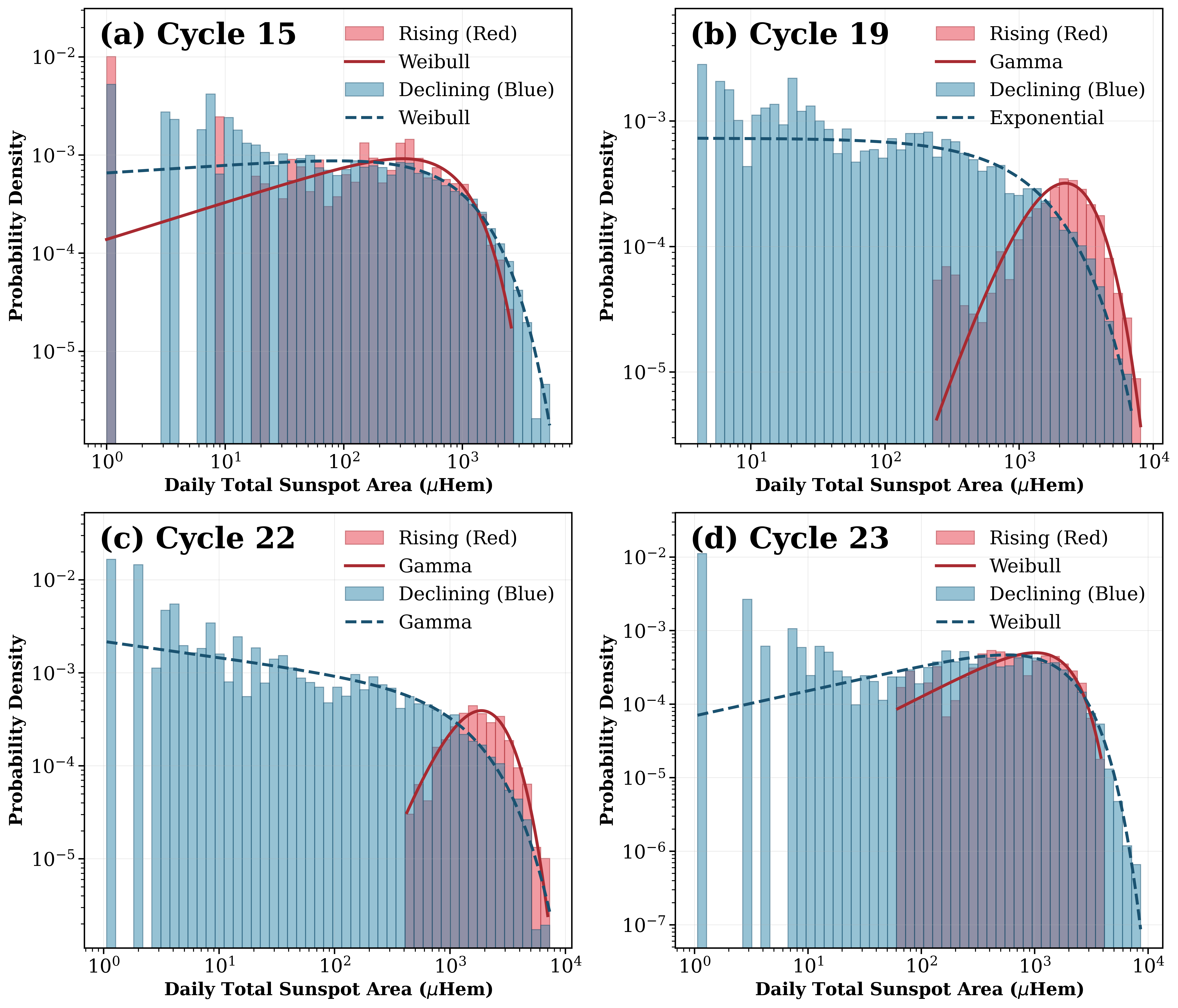}
\caption{Distributions of the daily group sunspot area (in \msh) during the rising (red) and declining (blue) phases of Cycles 15, 19, 22 and 23. 
}
\label{fig:distinvidual}
\end{figure*}

\subsection{Cross-checking results using RGO data}
\label{app:A2}
So far in the above analyses, we have used the sunspot area data from \citet{Mandal2020}, which does not keep track of the sunspot groups and thus the same group is recorded multiple times as independent observations. This means that if a group lives for a longer time (the big group), then it will be counted for a larger number than a shorter-lived group. To show that this weighting factor for the big groups does not affect our conclusion, we consider the sunspot data from Royal Observatory (RGO, Greenwich - USAF/NOAA), where the group numbers are recorded. By considering each sunspot group only once, we repeat our analyses. We find that the results remain qualitatively similar, and our conclusion remains unaltered. 
%As a demonstration, \Fig{fig:RGO} shows that the mean and median sunspot areas during the last three years in each cycle remain independent of the cycle strength. 

%\begin{figure*}%[H]
%\centering
%\includegraphics[width=0.96\textwidth]{corr_last3yr_str_rgo.png}
%\caption{The same as \Fig{fig:corr3yr} but produced from RGO data.}
%\label{fig:RGO}
%\end{figure*}

% Don't change these lines
\bsp	% typesetting comment
\label{lastpage}
\end{document}